\documentclass[aps,pre,twocolumn,superscriptaddress,longbibliography]{revtex4-2}
\usepackage{graphicx}
\usepackage{latexsym}
\usepackage{amssymb}
\usepackage{amsmath}
\usepackage{amsfonts}
\usepackage{upgreek}
\usepackage{float}
\usepackage{bm}
\usepackage{multirow}
\usepackage{color}
\usepackage[T1]{fontenc}
\usepackage{hyperref}
\usepackage{subfigure}
\usepackage{booktabs,tabularx}
\usepackage{bbm}

\hypersetup{
colorlinks = true,
linkcolor = [rgb]{0.70,0.13,0.13},
citecolor = [rgb]{0.13,0.55,0.13},
urlcolor  = [rgb]{0.25, 0.41, 0.88}}
\newcommand{\ket}[1]{|#1\rangle}

\begin{document}

\title{Non-Hermitian symmetry breaking and Lee-Yang theory for quantum XYZ and clock models}

\author{Tian-Yi Gu}
\affiliation{College of Physics, Nanjing University of Aeronautics and Astronautics, Nanjing, 211106, China}
\affiliation{Key Laboratory of Aerospace Information Materials and Physics (NUAA), MIIT, Nanjing, 211106, China}

\author{Gaoyong Sun}
\thanks{Corresponding author: gysun@nuaa.edu.cn}
\affiliation{College of Physics, Nanjing University of Aeronautics and Astronautics, Nanjing, 211106, China}
\affiliation{Key Laboratory of Aerospace Information Materials and Physics (NUAA), MIIT, Nanjing, 211106, China}

\begin{abstract}
Lee-Yang theory offers a unifying framework for understanding classical phase transitions and dynamical quantum phase transitions through the analysis of partition functions and Loschmidt echoes. Recently, this framework is extended to characterize quantum phase transitions of quantum Ising models by introducing the concepts of non-Hermitian parity-symmetry breaking and fidelity zeros. Here, we generalize the theory by studying a broad class of quantum models, including the XY, the XXZ, the XYZ, and the $\mathbb{Z}_p$ clock models in one dimension, subject to a complex magnetic field. For the XY, XXZ and XYZ models, we find that the complex field breaks parity symmetry and induces oscillations of the ground state between the two parity sectors, giving rise to fidelity zeros within the ordered phases. For the $\mathbb{Z}_3$ clock model, the complex field splits the real part of the ground-state energy between the neutral sector ($q=0$) and the charged sectors ($q=1,2$), while preserving the degeneracy within the charged sector. Fidelity zeros arise only after projecting out one of the charged sectors. For the $\mathbb{Z}_4$ and $\mathbb{Z}_5$ clock models, the ground states are instead projected to oscillate between the neutral sector ($q=0$) and the charged sectors $q=2$ and $q=1$, respectively, giving rise to fidelity zeros. Finite-size scaling of these zeros yields critical exponents in full agreement with analytical predictions, demonstrating that this approach is applicable not only to the Ising model with $\mathbb{Z}_2$ symmetry, but also to more general Heisenberg-type models and systems with higher discrete symmetries.

\end{abstract}

\maketitle

\section{Introduction}
Understanding phase transitions is one of the central topics in physics \cite{sachdev1999quantum}. The logarithm of the partition function, which is equivalent to the Helmholtz free energy in the canonical ensemble, serves as the central thermodynamic potential, with its nonanalytic behavior marking the emergence of a phase transition \cite{bena2005statistical,wang2025extracting}. However, because the partition function is composed of statistical weights that are analytic functions, its logarithm cannot develop any nonanalyticity in a finite system \cite{bena2005statistical}. To understand how phase transitions actually occur, Lee and Yang studied the partition function in finite systems under a complex magnetic field and developed a framework \cite{yang1952statistical,lee1952statistical} for characterizing phase transitions by analyzing its zeros, now known as Lee-Yang zeros \cite{lee1952statistical}. For the ferromagnetic Ising model, Lee and Yang proved that all zeros lie precisely on the unit circle \cite{lee1952statistical,bena2005statistical,wei2012lee}.

This compelling analysis of phase transitions and the distribution of zeros has attracted both theoretical \cite{bena2005statistical,heyl2018dynamical} and experimental \cite{peng2015experimental,brandner2017experimental} interest over the past decades.
The beauty of the Lee-Yang theory lies in identifying the zeros of the partition function in finite systems under a complex magnetic field, which appear only in systems with a real field in the thermodynamic limit. Fisher later extended this idea to study phase transitions by analyzing the zeros of the partition function (Fisher zeros) in the complex temperature plane \cite{fisher1965statistical}. Indeed, the key idea of analyzing the zeros of the partition function under a complex control parameter is quite general. For example, beyond the Lee-Yang zeros and Fisher zeros, it has been shown that phase transitions can be explored through complex $q$ in the $q$-state Potts model, giving rise to the concept of Potts zeros \cite{kim2001density}. Moreover, for quantum systems subjected to a sudden quench, the Loschmidt echo, which is analogous to the partition function upon defining the imaginary time $\tau = i t$ \cite{heyl2013dynamical}, is introduced to probe the system's dynamical behavior. The nonequilibrium phenomenon known as the dynamical quantum phase transition (DQPT) \cite{heyl2013dynamical} arises in the time plane at the zeros of the Loschmidt echo.

The distribution of zeros is another research interest. Originally, the Lee-Yang unit circle theorem was discovered in the ferromagnetic Ising model \cite{lee1952statistical}. It was subsequently extended to general Heisenberg spin models, showing that the theorem holds as long as the $z$-$z$ coupling is ferromagnetic and dominant \cite{suzuki1969distribution,asano1970lee,asano1970theorems,suzuki1971zeros}. The distribution of zeros on the unit circle implies that they concentrate along the imaginary axis of the magnetic field. Furthermore, it is shown that at high temperatures $T > T_c$, a gap develops in the distribution, giving rise to branch points, referred to as the Yang-Lee edge singularity \cite{fisher1978yang,kurtze1979yang,cardy1985conformal,cardy1989s,wei2017probing}. With decreasing temperature, the zeros approach and pinch the real axis \cite{ananikian2015imaginary}, signaling a phase transition at $T = T_c$ in the thermodynamic limit. Recently, Lee-Yang zeros and the Yang-Lee edge singularity have been revisited both theoretically \cite{yin2017kibble,ye2021dynamic,jian2021yang,sanno2022engineering,shen2023proposal,ye2023dynamic,ouyang2024complex,lu2025dynamical,xu2025characterizing,zhang2025yang,sun2025hybrid,xu2026thermodynamic} and experimentally \cite{gao2024experimental,lan2024experimental}.

In addition to classical and dynamical quantum phase transitions, quantum phase transitions \cite{sachdev1999quantum,sondhi1997continuous,vojta2003quantum} driven by quantum fluctuations are a central topic of broad interest in condensed matter physics. A natural extension of the Lee-Yang framework to quantum systems is to compute the partition function (or generating functions) of quantum many-body systems and analyze the distribution of zeros \cite{tong2006lee,kist2021lee,vecsei2022lee,vecsei2023lee,liu2024imaginary,liu2024exact,vecsei2025lee,li2023yang,li2025yang,meng2025detecting,wang2024quantum,he2025revisiting,lv2026giant}. Nevertheless, it is well-known that the most direct approach to quantum phase transitions is through the ground state and its associated observables. A key question is whether the Lee-Yang framework can be captured from the ground states of quantum many-body systems. This question has been addressed in recent work \cite{gu2026fidelity} based on the concepts of non-Hermitian symmetry breaking and fidelity (overlap of two ground states) zeros. 

As DQPTs may occur when a quench crosses a phase boundary but are absent for quenches within the same phase, they can serve as dynamical signatures of equilibrium phase transitions and therefore appear closely related to the fidelity-zero framework. However, we argue that the two frameworks differ in several key aspects. First, DQPTs typically require a large quench \cite{heyl2013dynamical}, whereas fidelity zeros arise from infinitesimal quenches and the corresponding ground-state overlaps. Second, DQPTs originate from dynamical orthogonality, whereas fidelity zeros arise from orthogonal ground states in different symmetry sectors. Third, DQPTs may occur for quenches within a phase and may be absent for quenches across a phase transition \cite{vajna2014disentangling,sedlmayr2018bulk}, whereas complex fields robustly switch between degenerate ground states at the transition. Finally, DQPTs may exhibit identical behavior across different transitions \cite{sun2020dynamical} and fail to capture $\nu = 5/6$ \cite{karrasch2017dynamical}, while fidelity zeros yield accurate critical exponents.

Fidelity zeros, examined in one- and two-dimensional ferromagnetic Ising models \cite{gu2026fidelity}, confirm the validity of the Lee-Yang theory for quantum phase transitions via the ground states in the complex magnetic field plane. However, an important open question remains: does this fidelity-based Lee-Yang framework apply only to the ferromagnetic Ising model, or can it be extended to a broader class of quantum many-body systems, analogous to the generality of the original Lee-Yang theory for classical phase transitions? In this work, we extend the framework developed in Ref. \cite{gu2026fidelity} to general quantum XYZ and $\mathbb{Z}_p$ clock models. For the XYZ model, we show that the Lee-Yang theory holds by examining the XY, XXZ and XYZ models, where parity is switched by the complex field, as in the Ising model.  In the $\mathbb{Z}_p$ clock model, the complex field similarly induces switching among the $\mathbb{Z}_p$ symmetry sectors. The fidelity-based Lee-Yang theory is formulated by projecting out selected symmetry sectors. This work therefore generalizes the fidelity-based Lee-Yang theory to broader classes of models and higher discrete symmetries, demonstrating the wide applicability of the Lee-Yang framework.

\begin{figure}[t]
\includegraphics[width=8.7cm]{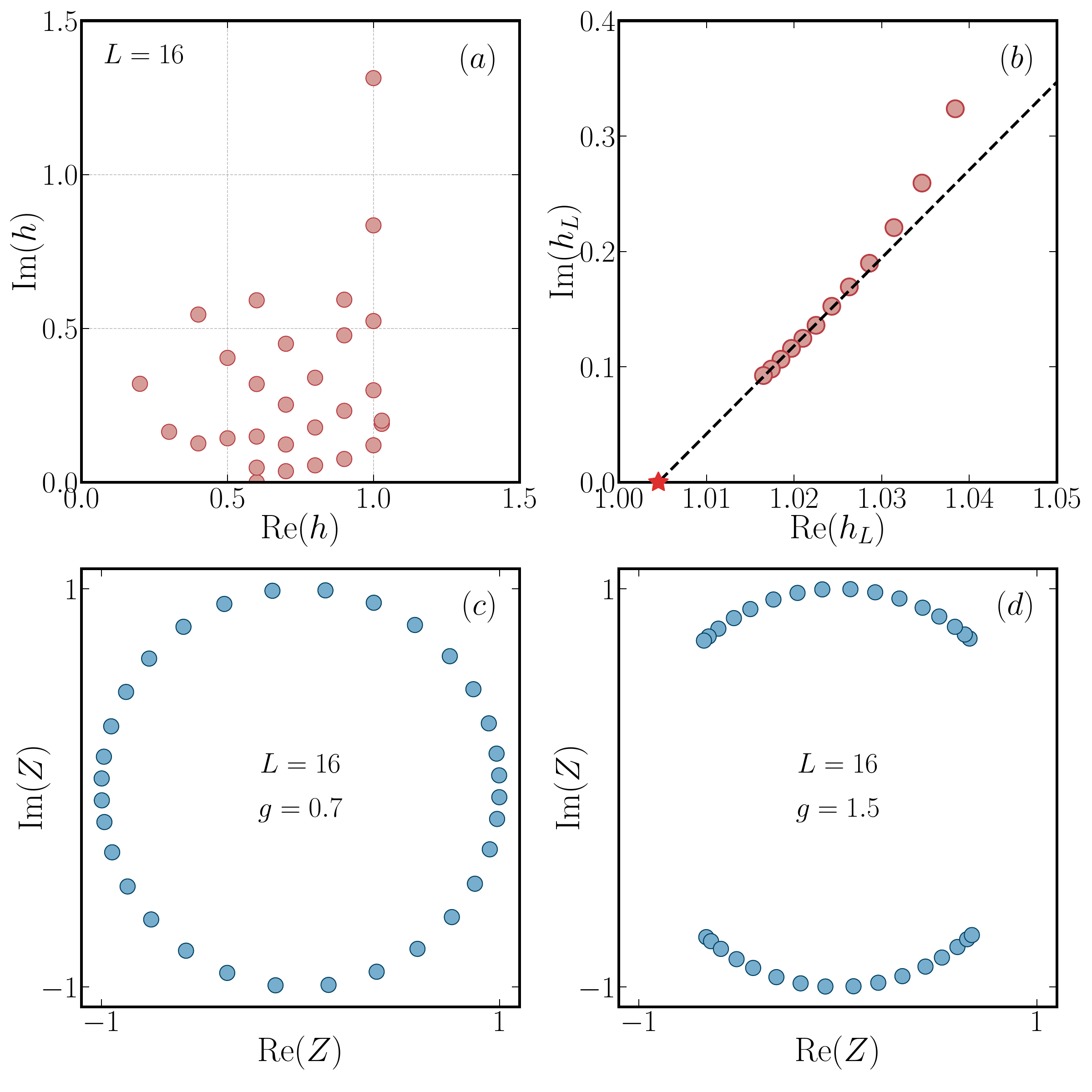} 
\centering
\caption{Fidelity zeros and fidelity edges in the XY model at $\gamma=0.8$. (a) Distribution of fidelity zeros in the complex-$h$ plane for $L = 16$. Scatter points indicate the complex field values where the fidelity vanishes, with the critical point located at $\mathrm{Re}(h) \approx 1$. (b) Finite-size scaling of the fidelity zeros as a function of system size $L$ ranging from 10 to 32. Dark-red dots represent the complex field $h$ with the maximum real and minimum imaginary parts; the black dashed line shows the fitted curve, and the red star on the real axis marks the critical value $h_c = 1.005$. (c) and (d) Distributions of fidelity zeros in the complex field plane, expressed as $h = g e^{i\theta}$, with $L = 16$ for $g = 0.7$ and $g = 1.5$, respectively. All zeros lie on the unit circle $Z = e^{i\theta}$ with $\theta \in (0, 2\pi]$, and fidelity edges appear for $g = 1.5$. }
\label{Fig:xy}
\end{figure}

\section{XYZ model} 
The general anisotropic Heisenberg Hamiltonian for the XYZ model with transverse and longitudinal magnetic fields is given by \cite{asano1970lee,asano1970theorems,suzuki1971zeros},
\begin{align}
H =& -\sum_{\langle i, j \rangle} \left( J_x S_{i}^{x} S_{j}^{x}  + J_y S_{i}^{y} S_{j}^{y} + J_z S_{i}^{z} S_{j}^{z} \right) \nonumber \\
      & \hspace{3.5cm} - \sum_{i} h_x S_{i}^{x} - \sum_{i} h_z S_{i}^{z},
\label{xyz_Ham}
\end{align}
where $J_x$, $J_y$ and $J_z$ denote the real anisotropic couplings along the $x$, $y$ and $z$ directions, respectively, and $h_x$ and $h_z$ represent the complex transverse magnetic fields, defined as either $h_{x,z}=h_{r} + i h_{i}$ or $h_{x,z}=ge^{i \theta}$ in the $x$ and $z$ directions.
Here, $S_{i}^{k}$ ($k=x, y, z$) denotes the spin operators acting on site $i$, and the notation $\langle i, j \rangle$ specifies a summation over all nearest-neighbor pairs in the lattice.
Throughout this work, periodic boundary conditions (PBCs) are imposed via $S_{L+1}^{x,y,z} = S_{1}^{x,y,z}$, with $L$ denoting the total number of spins in the chain.

\subsection{XY model}
We first consider the ferromagnetic XY model under a complex transverse field applied along the $z$ direction, with $J_x, J_y >0$, $J_z=0$ and $h_x=0$.
The Hamiltonian is given by
\begin{align}
H = -\sum_{\langle i, j \rangle} ( J_x S_{i}^{x} S_{j}^{x}  + J_y S_{i}^{y} S_{j}^{y} ) - \sum_{i} h_z S_{i}^{z},
\label{xy_Hamspin}
\end{align}
By parametrizing $J_x=2(1+\gamma)J$, $J_y=2(1-\gamma)J$ and $h=h_z/2$, and introducing the Pauli matrices $\sigma_{i}^{k} = 2S_{i}^{k}$ ($k=x, y, z$),
the Hamiltonian in Eq.~\eqref{xy_Hamspin} reduces to the anisotropic XY model given in \cite{pi2021phase}
\begin{equation}
H = - J \sum_{\langle i, j \rangle} \left( \frac{1+\gamma}{2} \sigma_{i}^{x} \sigma_{j}^{x} + \frac{1-\gamma}{2} \sigma_{i}^{y} \sigma_{j}^{y} \right)  - h \sum_{i} \sigma_{i}^{z},
\label{xy_Ham}
\end{equation}
where $J=(J_x + J_y)/4$, $\gamma = (J_x - J_y)/(J_x + J_y)$.
When $\text{Im}(h) = 0$, the Hamiltonian describes the Hermitian quantum XY model. In this case, $ J > 0$  indicates a ferromagnetic ordered phase, while $0 \leq \gamma \leq 1$ specifies the anisotropy strength. In the Hermitian regime, the XY model undergoes a quantum phase transition at $\text{Re}(h) = 1$. For $\gamma =1$, it reduces to the Ising model. In contrast, when $\text{Im}(h) \neq 0$, the Hamiltonian becomes non-Hermitian.

The non-Hermitian XY model can be solved exactly by mapping spin operators onto spinless fermions \cite{pi2021phase}.
To this end, it is convenient to rewrite Hamiltonian \eqref{xy_Ham} using the spin ladder operators $\sigma_{i}^{\pm} = \frac{1}{2}(\sigma_{i}^{x} \pm i\sigma_{i}^{y})$.
The Hamiltonian \eqref{xy_Ham} then becomes,
\begin{equation}
    \begin{split}
    	H =& - \sum_{i=1}^{L} [\sigma_{i}^{+} \sigma_{i+1}^{-} + \sigma_{i}^{-} \sigma_{i+1}^{+} + \gamma(\sigma_{i}^{+} \sigma_{i+1}^{+} + \sigma_{i}^{-} \sigma_{i+1}^{-})] \\
	    & - h \sum_{i=1}^{L} (2\sigma_{i}^{+} \sigma_{i}^{-} - 1).
    \end{split}
\label{xy_spinladder}
\end{equation}
The Jordan-Wigner transformation, which maps spin-1/2 particles to spinless fermions, yields the following relations,
\begin{equation}
	\begin{split}
		&\sigma_{i}^{+} = c_{i}^{\dagger} \exp\left[i\pi \sum_{j=1}^{i-1} c_{j}^{\dagger} c_{j}\right]  = c_{i}^{\dagger} \left( \prod_{j=1}^{i-1} \sigma_{j}^{z} \right) , \\
		&\sigma_{i}^{-} = \exp\left[-i\pi \sum_{j=1}^{i-1} c_{j}^{\dagger} c_{j}\right] c_{i} = \left( \prod_{j=1}^{i-1} \sigma_{j}^{z} \right) c_{i}, \\
		&\sigma_{i}^{z} = 2c_{i}^{\dagger} c_{i} - 1,
	\end{split}
\end{equation}
where $c_{i}^{\dagger}$ and $c_{i}$ are fermionic creation and annihilation operators satisfying the anticommutation relations $\{ c_{i}^{+},c_{j} \} = \delta_{ij} $ and $\{ c_{i},c_{j} \} = 0 $.
It is important to note that this transformation maps the original periodic boundary conditions (PBCs) to either PBCs or antiperiodic boundary conditions, depending on whether the system contains an odd or even number of particles. In the following analysis, we consider an even number of particles, giving the Hamiltonian,
\begin{align}
H_{\text{JW}} = & -\sum_{i=1}^{L} \left[ \left( c_{i}^{\dagger} c_{i+1} + c_{i+1}^{\dagger} c_{i} \right) + \gamma \left( c_{i}^{\dagger} c_{i+1}^{\dagger} + c_{i+1} c_{i} \right) \right] \nonumber \\
		&+ 2h \sum_{i=1}^{L} c_{i}^{\dagger} c_{i}.
\label{xy_JW}
\end{align}
Applying the Fourier transformation,
\begin{equation}
	c_{j} = \frac{1}{\sqrt{L}} \sum_{k} b_{k} e^{ikj},~~~ c_{j}^{\dagger} = \frac{1}{\sqrt{L}} \sum_{k} b_{k}^{\dagger} e^{-ikj},
\end{equation}
the Hamiltonian becomes,
\begin{equation}
H_{\text{FT}} = -\sum_k \left[ 2(\cos k - h) b_{k}^{\dagger} b_{k} + i \gamma \sin k \left( b_{k}^{\dagger} b_{-k}^{\dagger} + b_{k} b_{-k} \right) \right] 
\label{xy_FT}
\end{equation}
with $k$ taking the values $k = 2n\pi/L$ for $n = -L/2 + 1 , ... , L/2$.
By further applying the Bogoliubov transformation,
\begin{equation}
	\begin{split}
		a_k &= \cos{\theta_{k}} b_k - i \sin{\theta_{k}} b_{-k}^\dagger, \\
		a_{-k}^\dagger &= \cos^{\ast}{\theta_{k}} b_{-k}^\dagger - i \sin^{\ast}{\theta_{k}} b_k,
	\end{split}
\end{equation}
with the Bogoliubov angle $\theta_{k} $ defined by $\cos 2\theta_{k} = (h - \cos k)/ \sqrt{ (\cos k - h)^2 + \gamma^2 \sin^2 k} $, the Hamiltonian \eqref{xy_FT} is diagonalized as, 
\begin{equation}
	H_{\text{BG}} = \sum_{k>0} 2\epsilon_{k} \left( a_{k}^{\dagger} a_{k} + a_{-k}^{\dagger} a_{-k}  - 1 \right).
\label{xy_bog}
\end{equation}
where the operators $ a_{k} $ and $ a_{k}^{\dagger} $ are fermionic in momentum space and satisfy the same anticommutation relations as $ b_{k} $ and $ b_{k}^{\dagger} $. 
The energy spectrum of the Hamiltonian is given by,
\begin{equation}
\epsilon_{k} =  \sqrt{ (\cos k - h)^2 + \gamma^2 \sin^2 k },
\end{equation}
with the ground state denoted as,
\begin{equation}
|\Psi_0(h)\rangle = \prod_{k>0} \left( \cos\theta_k + i \sin\theta_k c_{k}^{\dagger}c_{-k}^{\dagger}\right) \ket{0}.
\end{equation}
The fidelity between $h$ and $h^{\prime}=h+\delta h$, where $\delta h$ is a small parameter change, is given by \cite{gu2026fidelity},
\begin{equation}
F(h, h^{\prime}) = \prod_{k>0} |( \cos^{\ast}\theta_k^{\prime} \cos\theta_k + \sin^{\ast}\theta_k^{\prime} \sin\theta_k )|.
\end{equation}
It is worth noting that for an odd number of fermions, the ground-state energy and fidelity can be directly computed using the allowed $k$ values $ k=\pm(2n - 1)\pi/L$, for $n = 1,\dots,L/2$.

\begin{figure}[t]
\includegraphics[width=8.7cm]{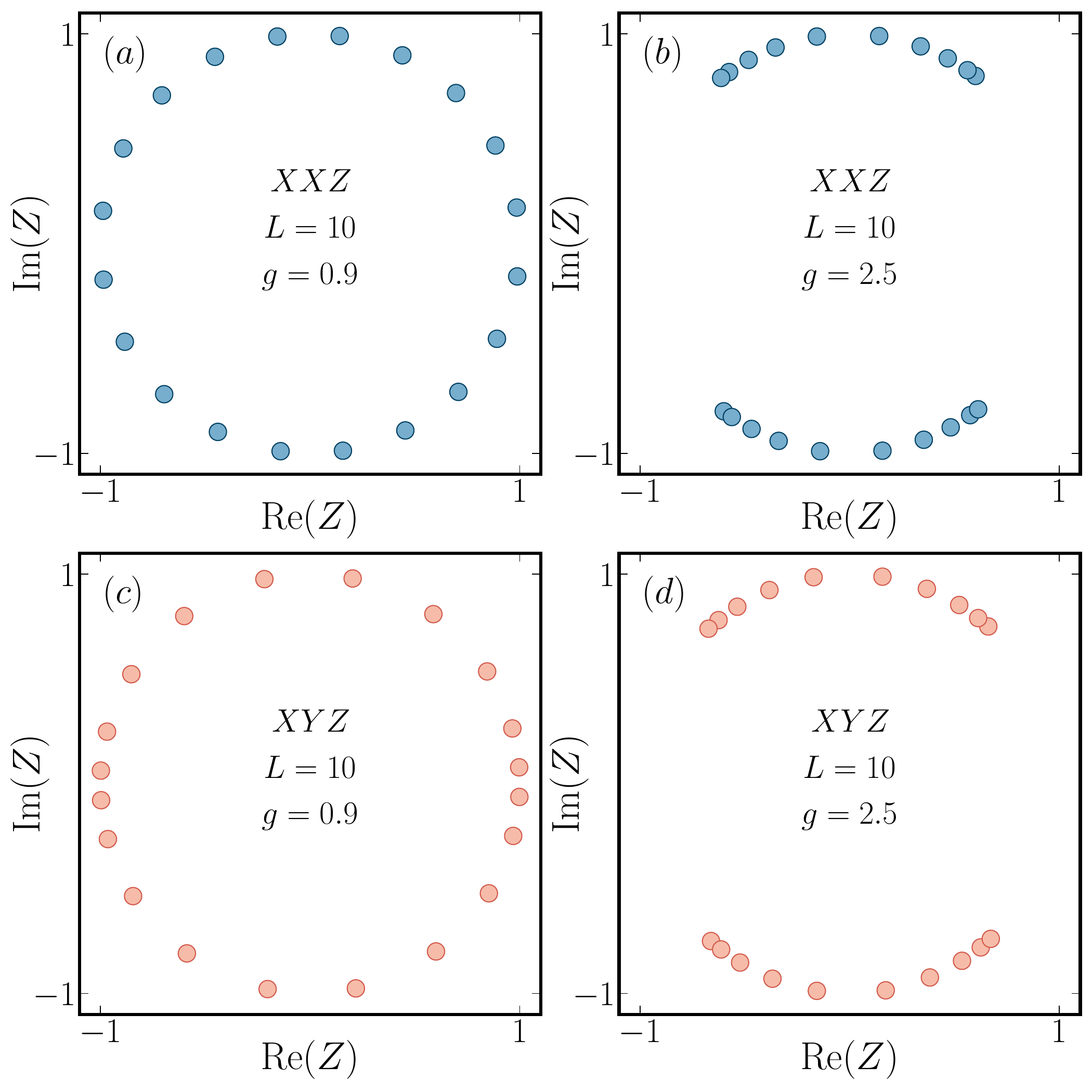} 
\centering
\caption{Distributions of fidelity zeros in the complex field plane for the XXZ and XYZ models, with $h = g e^{i\theta}$. (a)-(b) Fidelity zeros for the XXZ model with $L = 10$ for $g = 0.9$ and $g = 2.5$, respectively. (c)-(d) Fidelity zeros for the XYZ model with the same parameters as in (a)-(b). All zeros lie on the unit circle $Z = e^{i\theta}$ with $\theta \in (0, 2\pi]$, and fidelity edges appear for $g = 2.5$. }
\label{Fig:xyz}
\end{figure}

We analytically calculate the ground-state fidelity of the XY model by varying the real part of the complex magnetic field $h$ for $L = 16$ and $\gamma=0.8$, and show the corresponding distribution of fidelity zeros in Fig.~\ref{Fig:xy}, with $J=1$ fixed throughout the numerical simulations. As illustrated in Fig.~\ref{Fig:xy}(a), fidelity zeros appear in the system when $\text{Re}(h) < \text{Re}(h_{L})$, while no zeros are observed for $\text{Re}(h) > \text{Re}(h_{L})$. This behavior indicates that a quantum phase transition occurs at $\text{Re}(h_{L})$. 
To study the finite-size scaling of fidelity zeros near the critical point, we first define $h_{L}$ as the complex field $h$ with the largest real part and the smallest imaginary part in our finite-size analysis. The finite-size scaling of both $\text{Re}(h_{L})$ and $\text{Im}(h_{L})$ is analyzed using the relation \cite{gu2026fidelity},
\begin{equation}
	h_{L} = h_{c} + aL^{-1/ \nu },
	\label{eq:scaling}
\end{equation}
where $h_{c}$ and $a$ are fitting parameters, and $ \nu $ denotes the critical exponent of the correlation length.
The scaling behavior of fidelity zeros in the XY model is shown in Fig.~\ref{Fig:xy}(b) for system sizes $L$ ranging from 10 to 32. With increasing $L$, the critical points $h_{L}$ converge such that the real part approaches the real axis and the imaginary part vanishes in the thermodynamic limit, resulting in the critical point $h_{c} = 1.005$ and $ \nu = 1 $.

We further examine the distribution of fidelity zeros in the complex magnetic field $h = ge^{i \theta }$ for $L = 16$ [c.f. Fig.~\ref{Fig:xy}(c) and (d)].
These zeros are computed for $g = 0.7$ and $g = 1.5$, while $ \theta $ is varied over the interval $ (0,2 \pi] $.
By introducing the fugacity $Z = e^{i \theta }$, we observe that the zeros lie on the unit circle in the complex-$h$ plane.
For $g < h_{c}$, they are uniformly distributed along the circle; in contrast, for $g > h_{c}$, they condense into two circular arcs separated by a well-defined gap. These observations confirm that the Lee-Yang theory also holds for the XY model, as previously demonstrated for the transverse field Ising model \cite{gu2026fidelity}.

\begin{figure}[t]
\includegraphics[width=8.7cm]{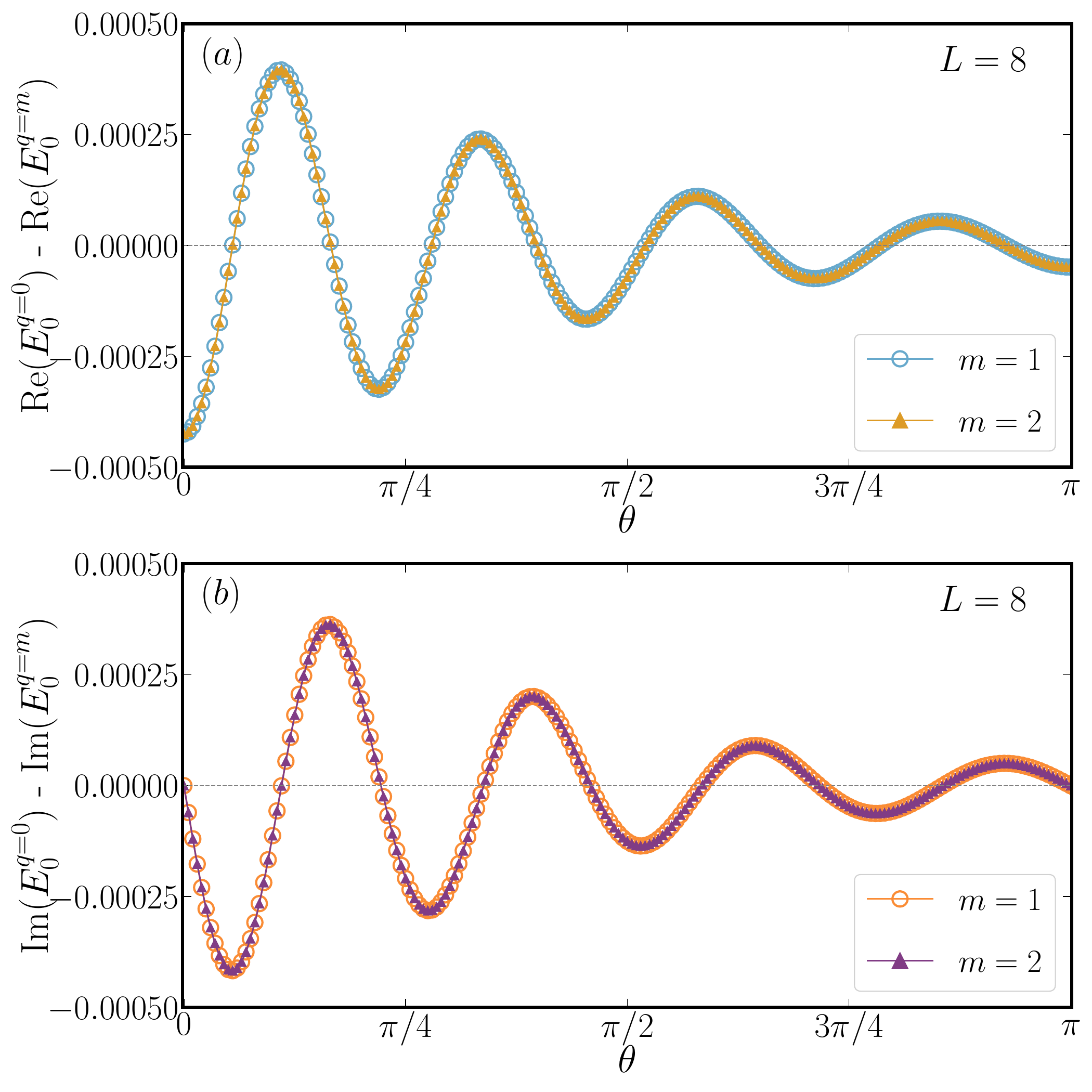} 
\centering
\caption{Ground-state energy differences of the clock model in different symmetry sectors for $L=8$ at $g=0.5$. (a) Real part of the ground-state energy difference, $\mathrm{Re}(E_{0}^{q=0}) - \mathrm{Re}(E_{0}^{q=1,2})$ between the $q=0$ and $q=1,2$ symmetry sectors as a function of $\theta$. (b) Imaginary part of the ground-state energy difference, $\mathrm{Im}(E_{0}^{q=0}) - \mathrm{Im}(E_{0}^{q=1,2})$ between the $q=0$ and $q=1,2$ symmetry sectors as a function of $\theta$. The numerical results are obtained under a complex transverse field $h = g e^{i\theta}$ with $\theta \in [0, \pi]$.}
\label{Fig:clock_energy}
\end{figure}

In the case $J_x =J_y \neq J_z$ and $h_z=0$, the system reduces to the XXZ model under a transverse field along the $x$ direction. When $J_x=J_y < 0$ and $J_z/J_x < -1$, the system undergoes a phase transition from a ferromagnetic to a paramagnetic phase \cite{dmitriev2002gap}. Furthermore, for the general XYZ model with $J_x  \neq J_y \neq J_z$ and $h_z=0$, a ferromagnetic-to-paramagnetic phase transition also occurs when the coupling $J_z > 0$ is dominant. Numerical simulations of the XXZ and XYZ models, presented in Fig.~\ref{Fig:xyz}, further support the validity of the Lee-Yang framework in characterizing quantum phase transitions in general spin systems.

\section{Clock model} 
We have shown that the fidelity-based Lee-Yang framework is valid for the general spin-1/2 XYZ model. In the following, we extend this theory to the $\mathbb{Z}_p$ clock model. The one-dimensional quantum $\mathbb{Z}_p$ clock model is described by the Hamiltonian~\cite{ortiz2012dualities,chen2017phase,sun2019phase,tang2023dynamical,yu2023dynamical,yu2024emergent}
\begin{equation}
	H = -J \sum_{j=1}^{L} \left(V_{j+1}^{\dagger} V_{j} + V_{j}^{\dagger} V_{j+1}\right) - h \sum_{j=1}^{L} \left(U_{j} + U_{j}^{\dagger}\right),
	\label{eq:q3clock}
\end{equation}
where $J$ denotes the coupling strength and and $h$ represents the complex transverse field.
The clock model can be regarded as a natural generalization of the Ising model and serves as a minimal model for investigating quantum systems with $\mathbb{Z}_p$ discrete symmetry. The local Hilbert space associated with the clock operators is $p$-dimensional, and the corresponding operators obey \cite{sun2019phase}
\begin{equation}
V_j U_l = e^{2\pi i \delta_{jl}/p} U_l V_j, 
\qquad 
U_j^p=V_j^p=\mathbbm{1}.
\end{equation}
In the orthonormal basis $\ket{s}$, with $s=0, 1, \cdots, p$, the clock matrices $V_j$ and $U_j$ are defined as \cite{sun2019phase},
\begin{equation}
V_j = \begin{pmatrix}
0 & 1 & 0 & \cdots & 0 \\
0 & 0 & 1 & \cdots & 0 \\
\vdots & \vdots & \vdots & \ddots & \vdots \\
0 & 0 & 0 & \cdots & 1 \\
1 & 0 & 0 & \cdots & 0
\end{pmatrix},
U_j = \begin{pmatrix}
1 & 0 & 0 & \cdots & 0 \\
0 & \omega & 0 & \cdots & 0 \\
0 & 0 & \omega^2 & \cdots & 0 \\
\vdots & \vdots & \vdots & \ddots & \vdots \\
0 & 0 & 0 & \cdots & \omega^{p-1}
\end{pmatrix},
\end{equation}
with $\omega=e^{2\pi i/p}$.
The diagonal clock operator $U_j$ acts as the phase operator, with eigenvalues $\omega^s$,
\begin{equation}
U_{j} \ket{s} = \omega^{s} \ket{s}.
\end{equation}
In contrast, the clock operator $V_j$ acts as a cyclic shift operator, 
\begin{equation}
V_{j} \ket{s} = \ket{s-1},
\end{equation}
where the index $s$ is defined modulo $p$. Thus,  $V_j$ and $V_j^\dagger$ serve as the cyclic lowering and raising operators, respectively.
Under the Fourier transformation of the local basis, the operator $V_j$ is transformed into a diagonal phase operator, whereas $U_j$ is transformed into a cyclic ladder operator. This interchange between $U_j$ and $V_j$ gives rise to the self-duality of the quantum clock model \cite{sun2019phase}. Consequently, the quantum clock model exhibits a second-order phase transition \cite{sun2019phase} at the self-dual point for $p\leq 4$, while for $p\geq 5$ it undergoes Berezinskii-Kosterlitz-Thouless (BKT) transitions. In the limit $p\rightarrow\infty$, the clock model reduces to the two-dimensional classical $XY$ model.

In the following, we investigate both second-order and BKT phase transitions using the fidelity-based Lee-Yang theory.
For $p=2$, the quantum clock model is exactly equivalent to the transverse-field Ising chain, which has been studied in Ref.\cite{gu2026fidelity}.
For $p=3$, the operators $V_{j}$ and $U_{j}$ are defined by,
\begin{equation}
	V_{j}=\begin{pmatrix}
		0 & 1 & 0 \\
		0 & 0 & 1 \\
		1 & 0 & 0
	\end{pmatrix},
	U_{j}=\begin{pmatrix}
		1 & 0 & 0 \\
		0 & \omega & 0 \\
		0 & 0 & \omega^{2}
	\end{pmatrix},
\end{equation}
respectively, where $\omega = e^{2\pi i/3}$ and $(V_{j})^{3}=(U_{j})^{3}=\mathbbm{1}$. Here, we impose $J=1$ and periodic boundary conditions $V_{L+1}=V_{1}$.
For a real field $\text{Im(h)}=0$, the $\mathbb{Z}_3$ clock model undergoes a phase transition at the critical point $\text{Re}(h)=1$, from a $\mathbb{Z}_3$-symmetric gapped ordered phase to a gapped disordered phase, with the correlation-length critical exponent $\nu=5/6$ \cite{sun2019phase}.

\begin{figure}[t]
\includegraphics[width=8.7cm]{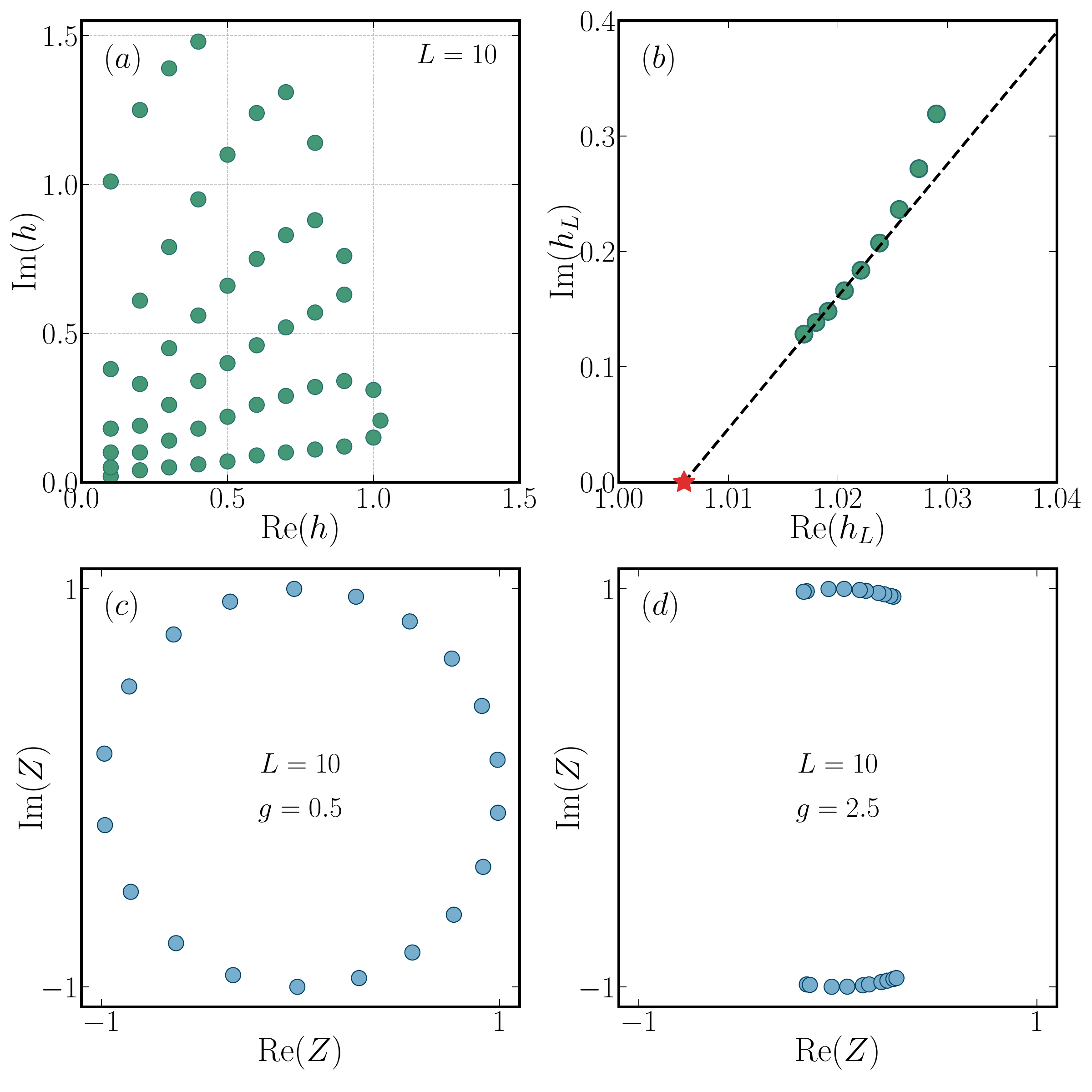} 
\centering
\caption{Fidelity zeros and fidelity edges in the $\mathbb{Z}_3$} clock model. (a) Distribution of fidelity zeros in the complex-$h$ plane for $L = 10$. The scatter points mark the complex field values at which the fidelity vanishes, with the critical point located at $\text{Re}(h) \approx 1$. (b) Finite-size scaling of the fidelity zeros as a function of system size $L$ from 7 to 15. The green dots denote the complex field values with the largest real part and smallest imaginary part of the fidelity zeros; the black dashed line shows the fitted curve, and the red star on the real axis marks the critical value $h_c = 1.006$. (c)-(d) Fidelity-zero distributions in the complex-field plane for $L = 10$ for $g = 0.5$ and $g = 2.5$, respectively. All zeros lie on the unit circle $Z = e^{i\theta}$ with $ \theta = (0,2\pi]$; Fidelity edges appear for $g = 2.5$.
\label{Fig:clock}
\end{figure}

In the ordered phase, the three degenerate ground states lie in distinct symmetry sectors, determined by the many-body $\mathbb{Z}_3$ projector,
\begin{equation}
	P_{q}= \frac{1}{3} \left[ \mathbbm{1} + \omega^{-q}\prod_{j} U_{j} + \omega^{-2q}\left(\prod_{j} U_{j}\right)^{2} \right],
\end{equation}
for the Hamiltonian labeled by $q=0,1,2$.
Specifically, $q=0$ identifies the neutral sector, whereas $q=1$ and $q=2$ belong to the charged sectors. 

However, when the external field is complex, for example $h=g e^{i\theta}$, it leads to the non-Hermitian symmetry-breaking and lifts the ground-state degeneracy, defined by the lowest real part of the complex energies. Notably, the complex field only lifts the degeneracy between the neutral sector ($q=0$) and the charged sectors ($q=1,2$), while the two ground-state energies within the charged sectors ($q=1,2$) remain degenerate, as shown in Fig.~\ref{Fig:clock_energy}. Because of this degeneracy, the ground state cannot be uniquely assigned to either the $q=1$ or $q=2$ sector.  Consequently, fidelity is ill-defined, as the ground state forms a superposition of states from the $q=1$ or $q=2$ sectors.
To lift this degeneracy and select a unique charged sector, we introduce an additional projector $P^{\prime} = \mathbbm{1} - P_{2} $, which completely projects out the $q=2$ sector. This results in a clear non-Hermitian symmetry-breaking between the $q=0$ and $q=1$ sector, which can be detected through fidelity zeros.

To validate our results, we numerically compute the fidelity zeros for a system of size $L=10$ in the complex field plane. 
The zeros are found exclusively within the ordered phase and vanish abruptly in the disordered phase, as shown in Fig.~\ref{Fig:clock}(a).
For $L=10$, the phase transition is located at $\text{Re}(h_L)=1.0238$. As the lattice size increases, the real part of $h$ approaches the real axis while the imaginary part vanishes in the thermodynamic limit, yielding a critical point $h_c = 1.006$ with $\nu = 5/6$, as illustrated in Fig.~\ref{Fig:clock}(b) through the finite-size scaling according to Eq. (\ref{eq:scaling}).
Furthermore, under the condition $h=gZ=ge^{i\theta}$, all fidelity zeros lie on the unit circle in the complex-$Z$ plane, with a total of $2L$ zeros. Their distribution, however, differs significantly between the ordered and disordered phases. For $g<h_c$, the zeros are distributed almost uniformly along the circle [c.f. Fig.~\ref{Fig:clock}(c)], whereas for $g<h_c$, they coalesce into two distinct arcs [c.f. Fig.~\ref{Fig:clock}(d)]. These results demonstrate that fidleity-based Lee-Yang approach is applicable to models with higher $\mathbb{Z}_3$ symmetries. 

For $p=4$ quantum clock model, the operators $V_j$ and $U_j$ are given by
\begin{align}
	V_{j}=\begin{pmatrix}
		0 & 1 & 0 & 0 \\
		0 & 0 & 1 & 0 \\
		0 & 0 & 0 & 1 \\
		1 & 0 & 0 & 0
	\end{pmatrix},
	U_{j}=\begin{pmatrix}
		1 & 0 & 0 & 0 \\
		0 & \omega & 0 & 0 \\
		0 & 0 & \omega^{2} & 0 \\
		0 & 0 & 0 & \omega^{3}
	\end{pmatrix},
\end{align}
respectively, where $\omega = e^{2\pi i/4} = i$ and $(V_{j})^{4}=(U_{j})^{4}=\mathbbm{1}$. For the $\mathbb{Z}_4$ clock model, the many-body projector can be constructed in close analogy to the $\mathbb{Z}_3$ case, and is explicitly expressed as
\begin{align}
	P_{q}= \frac{1}{4} \left[ \mathbbm{1} + \mathbb{T} +  \mathbb{T}^2 + \mathbb{T}^3 \right].
\end{align}
Here, $ \mathbb{T}=\omega^{-q}\prod_{j} U_{j}$, $q = 0$ is the neutral sector, and $q = 1, 2, 3$ are the charged sectors.

\begin{figure}[h]
	\includegraphics[width=8.7cm]{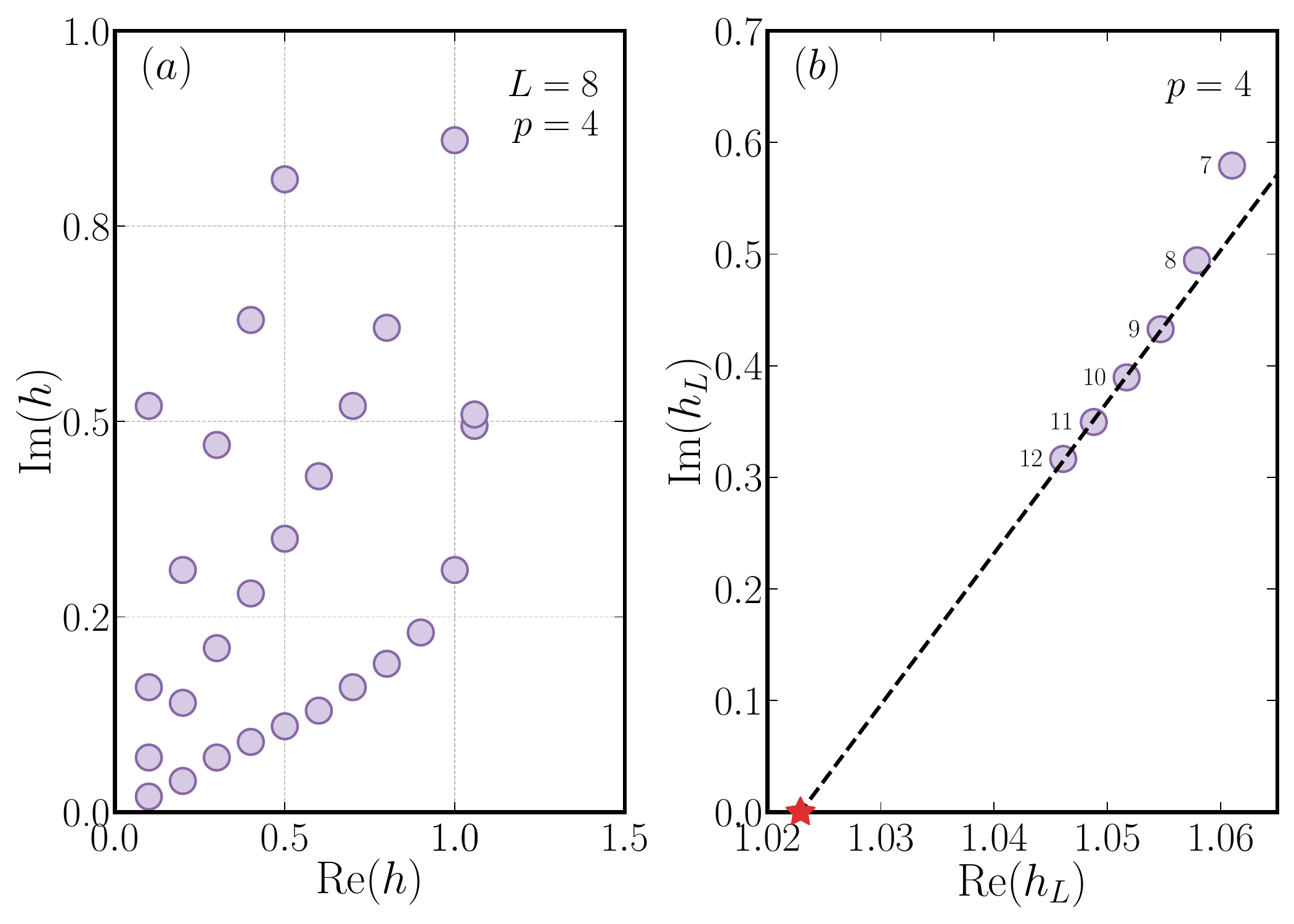} 
	\centering
	\caption{Fidelity zeros in the $\mathbb{Z}_4$ clock model. (a) Distribution of fidelity zeros in the complex-$h$ plane for $L = 8$. The scatter points mark the complex field values at which the fidelity vanishes, with the critical point located around $\text{Re}(h) \approx 1$. (b) Finite-size scaling of the fidelity zeros as a function of system size $L$ from $7$ to $12$. The purple dots denote the complex field values with the largest real part of the fidelity zeros; the black dashed line shows the fitted curve, and the red star on the real axis marks the critical value $h_c = 1.023$.}
	\label{Fig:q4clock}
\end{figure} 

For the $\mathbb{Z}_4$ clock model, the phase transition occurs at $\mathrm{Re}(h)=1$ with critical exponent $\nu=1$ under a real field \cite{sun2019phase}. In the presence of a complex field, non-Hermitian symmetry breaking lifts the ground-state degeneracy. Specifically, the complex field splits the spectrum among the neutral sector ($q=0$), the charged sector ($q=2$), and the pair of degenerate charged sectors ($q=1,3$). As a consequence, the ground state is effectively governed by the competition between the $q=0$ and $q=2$ sectors, giving rise to fidelity zeros that provide a clear signature of the phase transition.

We numerically calculate the fidelity zeros in the complex-field plane for a system of size $L=8$, as shown in Fig.~\ref{Fig:q4clock}(a). The zeros are prominently distributed around $\mathrm{Re}(h)=1$. For $L=8$, the finite-size transition point is found at $\mathrm{Re}(h_L)=1.0579$. Remarkably, fidelity zeros occur only in the region $\mathrm{Re}(h)<\mathrm{Re}(h_L)$, whereas they are absent for $\mathrm{Re}(h)>\mathrm{Re}(h_L)$. Applying the finite-size scaling analysis of Eq.~(\ref{eq:scaling}), we determine the critical point $h_c=1.023$ with the correlation-length critical exponent $\nu=1$, as illustrated in Fig.~\ref{Fig:q4clock}(b). These findings provide further evidence that the fidelity-zero framework can be systematically extended to models with higher discrete symmetries.

\begin{figure}[t]
	\includegraphics[width=8.3cm]{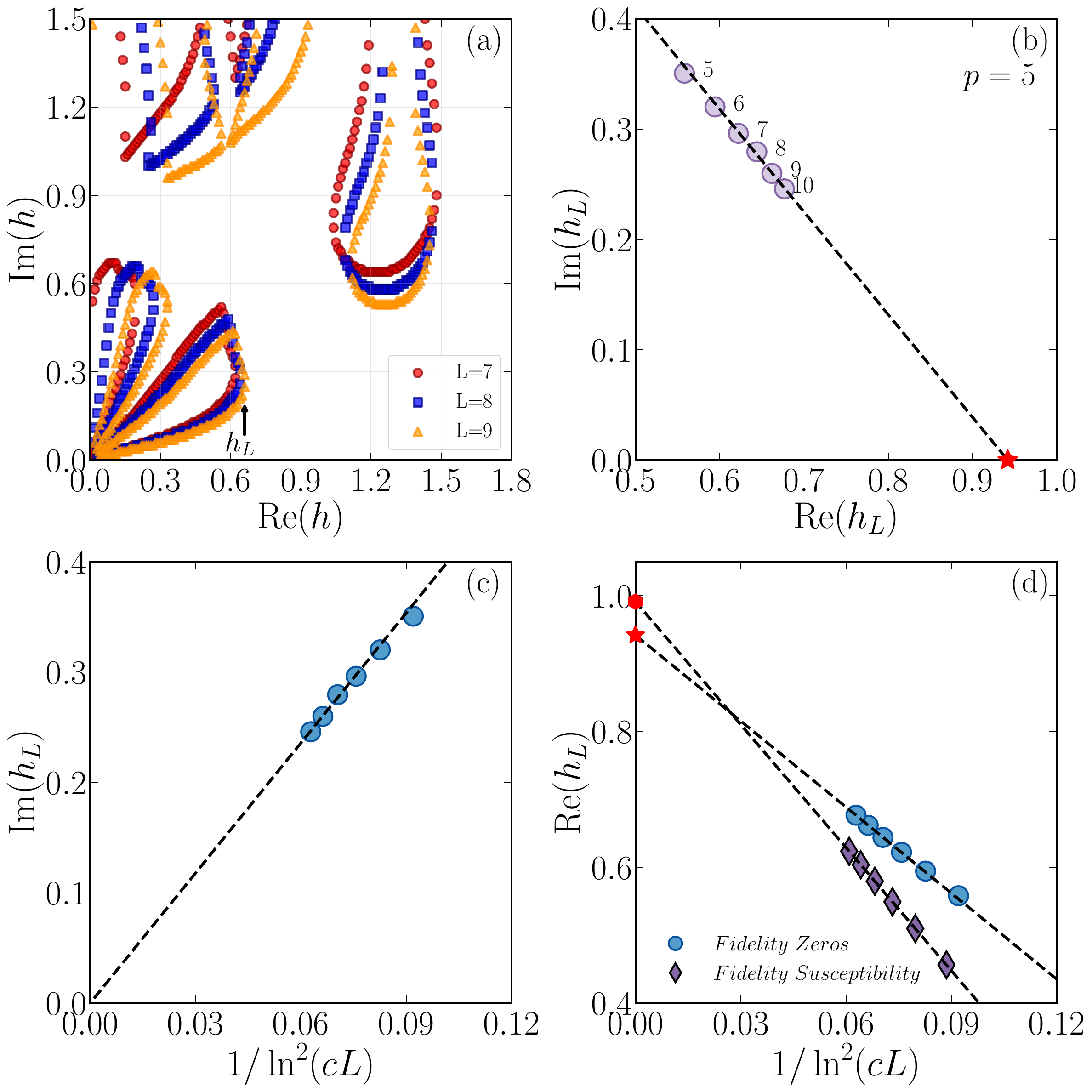} 
	\centering
	\caption{Fidelity zeros in the $\mathbb{Z}_5$ clock model in the $q=0$ and $q=1$ total charge sectors. (a) Distribution of fidelity zeros in the complex-$h$ plane for $L = 7, 8, 9$. The scatter points indicate the complex field values at which the fidelity vanishes. The black upward arrows ($\uparrow$) denote the finite-size pseudo-critical points $h_L$, the complex field values with the largest real part, used for finite-size scaling. (b) Finite-size scaling of the fidelity zeros with system size for $L=5$ to $10$. The purple dots indicate the finite-size pseudo-critical points $h_L$, the black dashed line represents the fitted curve, and the red star on the real axis marks the critical value $h_c = 0.9418$. (c) Finite-size scaling of the imaginary part of $h_L$, with $\mathrm{Im}(h_c)=0$. (d) The circle symbols denote the finite-size scaling of the real part of $h_L$, yielding $h_c=0.9420$. The diamond symbols denote the finite-size scaling of the pseudo-critical point $h_L$ obtained from the fidelity susceptibility (FS) in terms of the real field, giving $h_c^{\text{FS}}=0.9914$.}
	\label{Fig:q5clock}
\end{figure} 

For $p=5$ quantum clock model, the clock operators $V_j$ and $U_j$ are defined as
\begin{align}
	V_{j}=\begin{pmatrix}
		0 & 1 & 0 & 0 & 0\\
		0 & 0 & 1 & 0 & 0\\
		0 & 0 & 0 & 1 & 0\\
		0 & 0 & 0 & 0 & 1\\
		1 & 0 & 0 & 0 & 0 \\
	\end{pmatrix},
	U_{j}=\begin{pmatrix}
		1 & 0 & 0 & 0 & 0\\
		0 & \omega & 0 & 0 & 0\\
		0 & 0 & \omega^{2} & 0 & 0\\
		0 & 0 & 0 & \omega^{3} & 0 \\
		0 & 0 & 0 & 0 & \omega^{4} \\
	\end{pmatrix},
\end{align}
respectively, where $\omega = e^{2\pi i/5} = i$ and $(V_{j})^{5}=(U_{j})^{5}=\mathbbm{1}$. For the $\mathbb{Z}_5$ clock model, the many-body projector onto the symmetry sector labeled by $q$ can be written as
\begin{align}
	P_{q}= \frac{1}{5} \left[ \mathbbm{1} + \mathbb{T} +  \mathbb{T}^2 + \mathbb{T}^3 + \mathbb{T}^4 \right],
\end{align}
where $q=0$ corresponds to the neutral sector, while $q=1,2,3,4$ correspond to the charged sectors.
Owing to the charge-conjugation symmetry $q+q^{\prime}=p$ in the quantum clock model, the $q=1$ and $q=4$ sectors are degenerate in energy, as are the $q=2$ and $q=3$ sectors. Thus, in finite systems, there are three distinct states, corresponding to $q=0,1,2$, whose energies satisfy $E_0^{q=0} < E_0^{q=1} < E_0^{q=2}$.

To examine whether the Lee-Yang theory based on fidelity zeros can characterize the BKT transition, we compute the fidelity zeros of the $p=5$ clock model in the complex field $h$ and present the results in Fig.~\ref{Fig:q5clock}(a). We find that, similar to the cases with $p \leq 4$, fidelity zeros also emerge in the ordered phase at small complex $h$. We first determine the pseudo-critical points $h_L$ using the same procedure as that for the $p \leq 4$ cases, and then perform finite-size scaling of $h_L$, as shown in Fig.~\ref{Fig:q5clock}(b). The extrapolated critical point is $h_c=0.9418$, which is consistent with large-scale density-matrix renormalization group results ($h_c=0.966$, corresponding to $T_{\mathrm{BKT}}=0.913$ in the classical clock model \cite{sun2019phase}) and Monte Carlo simulations ($T_{\mathrm{BKT}}=0.908$, corresponding to $h_c=0.956$ in the quantum clock model \cite{kumano2013response}).

In addition, we carry out finite-size scaling analyses separately for the real and imaginary parts of the pseudo-critical points $h_L$. Both components are found to obey the logarithmic scaling form $h_L = h_c + a/\ln^2(cL)$, with $a$ and $c$ are fitting parameters, in agreement with the expected finite-size scaling behavior \cite{sun2019phase} of a BKT transition [c.f. Fig.~\ref{Fig:q5clock}(c) and (d)]. The critical point is estimated to be $h_c=0.9420$, consistent with the extrapolated value obtained from the complex-plane analysis, thereby further confirming the BKT nature of the transition. For comparison, we also determine the critical point from the fidelity susceptibility \cite{sun2019phase} with respect to a purely real field, obtaining $h_c^{\text{FS}}=0.9914$ [c.f. Fig.~\ref{Fig:q5clock}(d)]. These results indicate that our approach remains applicable to BKT transitions when degenerate ordered states are present, and may yield a more accurate estimate than the conventional fidelity-susceptibility method.

We note, however, that the $p=5$ clock model hosts two BKT transitions \cite{sun2019phase}. The first is the transition between the ordered phase and the gapless phase, which can be captured by our approach. The second occurs between the gapless phase and the disordered gapped phase. This transition cannot be described within our theory, since there are no degenerate ground states that can switch under a complex field. Nevertheless, fidelity zeros may still appear in this regime, as shown in Fig.~\ref{Fig:q5clock}(a). A complete understanding of these zeros lies beyond the scope of the present approach and of this paper, and is left for future study.

\section{Conclusion} 
In summary, we have generalized the Lee-Yang theory based on fidelity zeros to both the XYZ model and the $\mathbb{Z}_3$ clock model. The fidelity zeros in these systems, arising from non-Hermitian symmetry breaking, exhibit distinct distributions in the ordered and disordered phases, consistently obey the Lee-Yang theorem, and faithfully signal the phase transition. Our results show that this approach is not only effective for the anisotropic XYZ model but also capable of capturing criticality in systems with higher discrete symmetries, such as the $\mathbb{Z}_p$ clock model. 

Consequently, the fidelity-based Lee-Yang framework provides a simpler implementation with robust numerical accuracy for identifying quantum phase transitions, thereby opening avenues for applying the Lee-Yang formalism to a broader class of correlated quantum systems. A promising direction for future work is to explore its extension to continuous symmetry-breaking cases, such as U(1) and SU(2) symmetries.

\section{Acknowledgments} 
G.S. is appreciative of support from the NSFC under the Grants No. 11704186 and "the Fundamental Research Funds for the Central Universities, NO. NS2023055".
T.-Y. G acknowledges support from the Postgraduate Research \& Practice Innovation Program of NUAA under Grant No. xcxjh20252113. 
This work is partially supported by the High Performance Computing Platform of Nanjing University of Aeronautics and Astronautics.

\section*{Data Availability}
The data that support the findings of this article are not publicly available. The data are available from the authors upon reasonable request.

\bibliography{XYZCref}

\end{document}